\documentstyle[preprint,aps,epsfig]{revtex}

\def\be{\begin{equation}}
\def\ee{\end{equation}}

\def\ba{\begin{array}}
\def\ea{\end{array}}
\def\bea{\begin{eqnarray}}
\def\eea{\end{eqnarray}}

\def\l{\left}
\def\r{\right}

\begin{document}
%\draft 

\rightline{CERN-TH/2000-044}
\rightline{IMSc/2000/02/04}
\rightline{UMD-PP-00-054}
\rightline{hep-ph/0002177}

\begin{center}

{\large \bf Radiative magnification of neutrino mixings and a natural
explanation  of the neutrino anomalies}

\bigskip

{K. R. S. Balaji\footnote{balaji@imsc.ernet.in}}

{\it Institute of Mathematical Sciences, Chennai 600 113, India.}

\medskip

{Amol S. Dighe\footnote{Amol.Dighe@cern.ch}}

{\it Theory Division, CERN, CH-1211 Geneva 23, Switzerland.}

\medskip

{R. N. Mohapatra\footnote{rmohapat@physics.umd.edu}}

{\it Department of Physics, University of Maryland, 
College park MD 20742, USA.} 

\medskip

{M. K. Parida\footnote{mparida@vsnl.com}}

{\it Department of Physics, North Eastern Hill University, Shillong
793022, India.} 

\end{center}

%\date{\today}
%\maketitle
%\vspace{1.0cm}

\begin{abstract}

We show that the neutrino mixing pattern with the 
large mixing required for the atmospheric neutrino
problem and the small mixing angle MSW solution for the
solar neutrino problem can be naturally generated 
through radiative magnification, even though all
the mixing angles at the seesaw scale may be small.
This can account for the neutrino anomalies as well as
the CHOOZ constraints in the context of quark-lepton
unified theories, where the quark and lepton mixing 
angles are expected to be similar in magnitude 
at the high scale.
We also indicate the 4$\nu$ mixing scenarios for
which this mechanism of radiative magnification
can provide a natural explanation.

\end{abstract}

\vspace{2.0cm}
%\pacs{PACS numbers: 14.60.Pq, 12.15.Lk, 12.15.Ff, 12.60.-i} 
\leftline{PACS numbers: 14.60.Pq, 12.15.Lk, 12.15.Ff, 12.60.-i}

\newpage

\section{INTRODUCTION}
A major theoretical challenge posed by the solutions to the 
atmospheric and solar neutrino anomalies \cite{atm,solar}
is that the atmospheric neutrino data
require a large $\nu_{\mu}-\nu_{\tau}$ mixing, whereas the
corresponding quark mixing between the second
and the third generation is very small.
This is not easy to understand in the context of quark-lepton unified
theories. 
While there are suggestions to understand a large mixing
in the context of various kinds of unified
theories \cite{maximal} 
including the SO(10), where there is a natural
quark-lepton unification\cite{barr}, no 
convincing natural model has yet emerged. 
It is therefore necessary to explore
alternative possibilities. 
One way to proceed is not to concentrate on a particular
model, but to look for the features that a model should
have in order to be able to predict the observed large
mixing naturally.

In a recent paper \cite{balaji}, we pointed out 
that for two Majorana neutrinos with the same CP parity 
that are nearly degenerate in mass, a small 
neutrino mixing at the high scale 
can be magnified by the 
radiative corrections through the
renormalization group running down
to the weak scale. 
In such theories, there would be no need to put special 
constraints on the mixings in the theory at the 
high ({\it e.g.} seesaw) 
scale and indeed the quark and
lepton mixings can be very similar (as, say, 
would be predicted by the simple seesaw models). 
It is the goal of this paper to show that
such a mixing pattern, which involves only small mixings 
at the high scale ($\Lambda$), can explain the neutrino
anomalies at the low scale as long as the conditions outlined
in \cite{balaji} are satisfied. It is then possible to
explain the solar neutrino problem through the small angle
MSW solution and the atmospheric neutrino problem through
the large mixing angle, which gets generated through the
radiative magnification.

The paper is organized as follows. 
In Sec.~\ref{mag}, we 
introduce the $(\Omega,\Phi,\Psi)$ parametrization for the
mixing angles, and taking $\Phi=0$ at the high scale, show 
that the radiative corrections can magnify $\Psi$ while
keeping $\Omega$ and $\Phi$ unaffected. 
In Sec.~\ref{fit},
we show that the condition $\Phi=0$ is consistent with the current
data from the solar, atmospheric and reactor experiments.
In Sec.~\ref{4nuscheme}, we consider the possible 4$\nu$ mixing schemes
that can explain the LSND results in addition, and identify
the scenarios for which radiative magnification can provide
a natural explanation through the quark-lepton unified theories.
Sec.~\ref{concl} concludes.

\section{Radiative magnification for three neutrino mixing}
\label{mag}

In the absence of $CP$ violation in the lepton sector,
the mixing matrix $U_\Lambda$ at the scale $\Lambda$ 
can be parametrized as 
\be
U_\Lambda ~=~ 
U_{12}(\Omega) \times U_{13}(\Phi) \times U_{23}(\Psi)~~,
\label{split-u}
\ee
where all the three rotation angles lie between 0 and $\pi/2$.
Note that the order of multiplication of the rotation matrices
is different from the conventional one \cite{kp}, so the
angles $\Omega,\Phi,\Psi$ involved here should not be mistaken
for the angles $\omega,\phi,\psi$ used conventionally. 
Nevertheless, (\ref{split-u}) is a perfectly valid way of
parametrizing the mixing matrix, and is useful for addressing
a certain class of problems ({\it e.g.} see \cite{derujula}).

At the low scale $\mu$, the mixing matrix $U_\mu$ can be written
in general as
\be
U_\mu ~=~ 
U_{12}(\bar{\Omega}) \times U_{13}(\bar{\Phi}) 
\times U_{23}(\bar{\Psi})~~.
\label{split-ubar}
\ee
The CHOOZ results \cite{chooz} indicate a small 
$U_{e3}$, 
which corresponds to a small value for
$$
\cos \bar{\Omega} \cos \bar{\Psi} \sin \bar{\Phi} 
+ \sin \bar{\Omega} \sin \bar{\Psi}  ~~.
$$
This can be satisfied with the choice of $\bar{\Phi}=0$
and a small $\sin \bar{\Omega} \sin \bar{\Psi}$. That such a
choice can satisfy the solar and the atmospheric data
is shown in Sec.~\ref{fit}.
With this motivation, we start with $\Phi = 0$ at the
high scale (this choice leads to $\bar{\Phi}=0$, as we
shall show in this section), and show that the radiative
corrections can magnify $\Psi$ while keeping $\Omega$
and $\Phi$ unaffected.

With only the $\Omega$ and $\Psi$ mixings nonzero at the scale $\Lambda$,
the effective mass matrix $M^{eff}_{\Lambda}$ in the flavor
basis is
\be
M^{eff}_{\Lambda}~=
~ U_\Lambda ~M_\Lambda ^d ~ U_\Lambda^\dagger
=
~ U_{12}(\Omega) ~U_{23}(\Psi) ~M_\Lambda ^d 
~U_{23}^\dagger(\Psi) ~U_{12}^\dagger(\Omega)~~,
\label{m-lamb}
\ee
where $M_\Lambda^d = Diag(m_1, m_2, m_3)$.
If the radiative corrections are included \cite{babu,haba}, we
have
\be
M^{eff}_{\Lambda} \rightarrow ~M^{eff}_{\mu} ~=~
\l(\ba{ccc} \sqrt{I_e} & 0 & 0\\ 0 &\sqrt{I_\mu} & 0\\ 0 & 0 &
 \sqrt{I_\tau} \ea \r)
M^{eff}_{\Lambda}
\l(\ba{ccc} \sqrt{I_e} & 0 & 0\\ 0 &\sqrt{I_\mu} & 0\\ 0 & 0 &
\sqrt{I_\tau} \ea \r),
\label{matrices}
\ee 
where $I_\alpha -1 \equiv 2\delta_\alpha$ are the 
radiative corrections that appear due to the 
Yukawa couplings of the charged leptons $e, \mu$ and
$\tau$ respectively. Given the strong hierarchical pattern of the
charged lepton
masses, we neglect the corrections due to $e$ and $\mu$, 
{\it i.e.} $I_e = I_\mu=1$. 
Let us define ${\cal I}_\tau \equiv Diag(1,1,\sqrt{I_\tau})$.
Then from (\ref{m-lamb}) and (\ref{matrices}), 
\be
M^{eff}_{\mu}~=~ {\cal I}_\tau  ~U_{12}(\Omega)
~U_{23}(\Psi)  ~M_\Lambda^d 
~U_{23}^\dagger(\Psi)
~U_{12}^\dagger(\Omega) ~{\cal I}_\tau ~~.
\label{iuum}
\ee
Noting that $[U_{12}(\Omega),{\cal I}_\tau] = 0$, we get
\be
M^{eff}_{\mu}~=~ U_{12}(\Omega)
~[{\cal I}_\tau ~U_{23}(\Psi) ~M_\Lambda^d 
~U_{23}^\dagger(\Psi)
~{\cal I}_\tau] ~U_{12}^\dagger(\Omega)~~.
\label{uium}
\ee
The quantity in the square brackets in (\ref{uium}) is
in a form where the first row and column are
effectively decoupled and the situation reduces
to the two-generation mixing,
which has been considered in detail in \cite{balaji}.
This quantity can be written as
\be
{\cal I}_\tau ~U_{23}(\Psi) ~M_\Lambda^d ~U_{23}^\dagger(\Psi)
~{\cal I}_\tau =
U_{23} (\tilde\Psi) ~M_\mu^d ~U_{23}^\dagger(\tilde\Psi)~~,
\label{m-mu}
\ee
where $M_\mu^d$ is a diagonal matrix. The new (2-3) mixing
angle $\tilde{\Psi}$ is given by
\bea
\tan (2\tilde \Psi)&=& \frac{\tan(2 \Psi)}{\lambda}~
(1 + \delta \tau)~~, \nonumber\\
\lambda &\equiv& 
\frac{(m_3~-~m_2)C_{2\Psi}~+~2\delta_\tau m}{(m_3~-~m_2)C_{2\Psi}}~,
\label{tanpsi}
\eea
where $m$ is the common mass of the quasi-degenerate neutrinos. 
Now, if
\be
\delta_\tau ~\approx~ \frac{(m_2-m_3) C_{2\Psi}}{2m}~,
\label{cond}
\ee
then $\lambda \approx 0$, so that the mixing angle $\tilde{\Psi}$
becomes large \cite{balaji}. Since $\delta_\tau \ll 1$, for
the condition (\ref{cond}) to be satisfied, $\nu_2$ and $\nu_3$ 
need to have the same CP parity.
Thus the $\Psi$-mixing can be magnified at the weak scale, 
which explains the atmospheric neutrino data
(See Sec.~\ref{fit}).

From (\ref{uium}) and (\ref{m-mu}),
\be
M^{eff}_{\mu}~=~ U_{12}(\Omega)
~U_{23} (\tilde\Psi) ~M_\mu^d 
~U_{23}^\dagger(\tilde\Psi)
~U_{12}^\dagger(\Omega)~~.
\label{uum}
\ee
This shows that the same (1-2) mixing angle $\Omega$ 
that was needed for diagonalizing $M^{eff}_{\Lambda}$ 
is also needed for
diagonalizing $M^{eff}_{\mu}$ [see (\ref{m-lamb}) and (\ref{uum})], 
and that a (1-3) mixing angle $\Phi$ is not required. 
Thus, $\bar{\Psi} = \tilde{\Psi}$, $\bar{\Omega} = \Omega$ and
$\bar{\Phi} = \Phi = 0$ are the mixing angles at the low scale.

As we shall see in Sec.~\ref{fit}, we can explain the solar, 
atmospheric and the CHOOZ data with $\bar{\Psi} \approx \pi/4$
and a small $\bar{\Omega}$ (corresponding to the SMA solution
for the solar neutrinos).
In a typical quark-lepton unified theory,
$\Omega$ would be small at the high scale.
In the limit of neglecting the radiative corrections
due to the second generation 
({\it i.e.} $I_\mu \to 1$) that we have considered here,
the magnification of $\Omega$ due to radiative corrections
is not possible. Also, if the $CP$ parity of the neutrino
$\nu_1$ is opposite to that of $\nu_2$ and $\nu_3$
(which is required to ascertain the stability of
a possible small nonzero $\Phi$), a small
$\Omega$ at the high scale will stay small even when the radiative
corrections due to $\mu$ are taken into account. Thus, 
the stability of a small $\Omega$ is guaranteed, and
the small angle MSW scenario can be generated naturally
within the unification
models.

The radiative corrections from the second
generation [{\it i.e.} ${\cal I}_{\mu} \equiv Diag(1,\sqrt{I_\mu},1)
\neq Diag(1,1,1)$] modify 
(\ref{uium}) to
\be
M^{eff}_{\mu}~=~{\cal I}_{\mu} ~U_{12}(\Omega)
~[{\cal I}_\tau ~U_{23} (\Psi) ~M_\Lambda^d 
~U_{23}^\dagger (\Psi)
~{\cal I}_\tau] ~U_{12}^\dagger(\Omega)~{\cal I}_{\mu}~~.
\label{uium1}
\ee
Since $[U_{12}(\Omega), {\cal I}_\mu] \neq 0$, 
the value of $\Omega$ may now get modified and $\Phi$ may
get generated. The value of $\bar{\Psi}$ is also 
different from the value of $\tilde{\Psi}$ as given in 
(\ref{tanpsi}). But since 
\be
[U_{12}(\Omega), {\cal I}_\mu] = \delta_\mu ~\sin \Omega 
~ \l(\ba{cc} 0 & 1 \\ 1 & 0 \ea \r) ~~,
\ee
and the values of 
$\Omega$ and $\delta_\mu$ are both small, these differences
$(\bar{\Omega}-\Omega)$, $(\bar{\Phi}-\Phi)$ and
$(\bar{\Psi} - \tilde{\Psi})$ are not expected to be 
large.

\section{Satisfying the solar, atmospheric and CHOOZ data}
\label{fit}

In the following, we show that our choice of 
parametrization (\ref{split-u}) with  $\Phi=0$ 
can explain the solar and atmospheric 
anomalies and still
be consistent with the stringent bounds coming 
from the CHOOZ experiment.

We first concentrate on the CHOOZ and the atmospheric data 
which share a common mass scale 
$\Delta m^2_{31} \approx \Delta m^2_{32} \approx 10^{-3} eV^2$. 
In this case, the relevant probability
expressions are
\bea
P_{\alpha\beta}^{atm} & \approx &4~|U_{\alpha 3}|^2 |U_{\beta 3}|^2 
\sin^2 \left( \frac{\Delta m^2_{31} L}{4E} \right)~~,
\label{pab} \\
P_{\alpha\alpha}^{atm} &\approx & 
1~-~4|U_{\alpha 3}|^2( 1~-~|U_{\alpha 3}|^2)
\sin^2 \left( \frac{\Delta m^2_{31} L}{4E} \right)~.
\label{paa}
\eea
To satisfy the CHOOZ constraint \cite{chooz}
\be
|U_{e3}|^2 < 0.03 \quad \mbox{ for } 
\Delta m^2_{31} > 2 \cdot 10^{-3} \mbox{ eV}^2
\label{chooz-limit}
\ee
with $\bar{\Phi} = 0$, we need
\be
\sin \bar{\Omega} \sin \bar{\Psi} < 0.17~~.
\label{cond1}
\ee
This small value of $|U_{e3}|^2$ 
also guarantees $P_{ee} \approx 1$ [eq. (\ref{paa})]
and $P_{\mu e} \approx 0$ [eq. (\ref{pab})]
in the atmospheric neutrino data.

A fit to the $L/E$ distribution of the atmospheric neutrinos
\cite{l/e} gives at 90\% confidence level, using (\ref{paa}),
\be
0.2 < |U_{\mu 3}|^2 < 0.8~~.
\label{umu3}
\ee
This corresponds to
\be
0.45 < \cos \bar{\Omega} \sin \bar{\Psi}  <  0.9~~.
\label{cond2}
\ee

By examing the mixing matrix as parametrized in (\ref{split-ubar}),
we can see that
$\bar{\Phi} = 0$, $\bar{\Psi} \approx \pi/4$ and
a small $\bar{\Omega}$ can easily satisfy the requirements
(\ref{cond1}) and (\ref{cond2}). The smallness of 
$\bar{\Omega}$ is forced by the CHOOZ constraints and
the large value of $\sin \bar{\Psi}$ is required by the
atmospheric $P_{\mu \mu}$.

\begin{figure}
\vspace{1.0in}
\epsfig{file=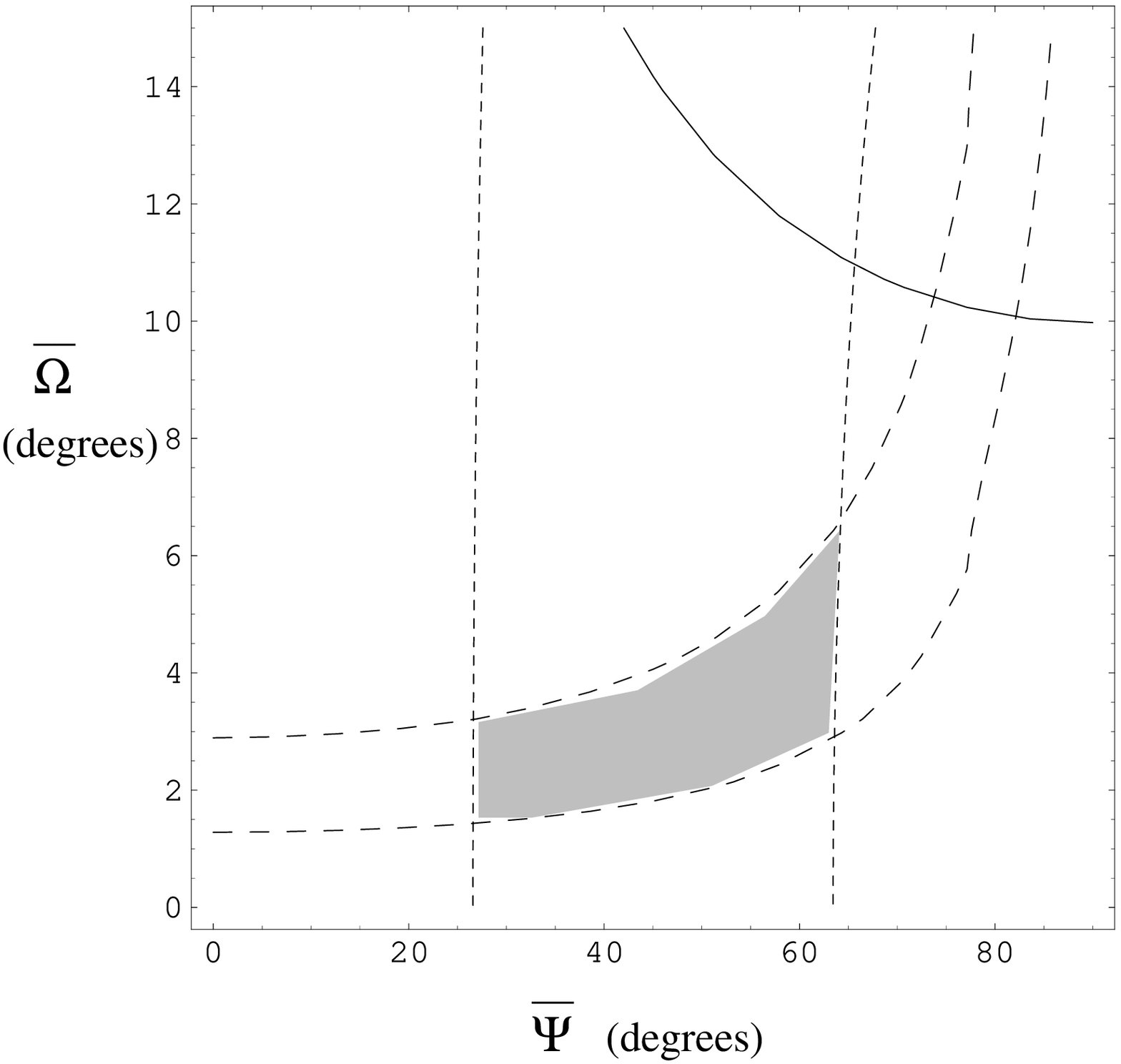,width=5in}
\vspace{0.5in}
\caption{The allowed parameter space for $\bar{\Omega}$ and 
$\bar{\Psi}$. The region below the solid line is allowed by
CHOOZ, the region between the almost vertical dotted lines is
allowed from the atmospheric data and the region between the
dashed lines corresponds to the SMA solution for the solar
neutrinos. The shaded region is consistent with all the data.}
\label{param-space}
\end{figure}

Let us now consider the solar neutrino solution.
In the case of the solar neutrino anomaly, the
SMA solution corresponds to 
\be
|U_{e2}|^2 \approx (0.5 \div 2.5) \cdot 10^{-3}~~,
\label{ue2-limit}
\ee
whereas the other solutions -- LMA, LOW and VO --
correspond to $|U_{e2}|^2 > 0.2$. 
In the parametrization (\ref{split-ubar}),
\be
|U_{e2}|^2 = \sin^2 \bar{\Omega} \cos^2 \bar{\Psi}~~.
\label{ue2new}
\ee
It is difficult to reconcile the smallness of 
$\bar{\Omega}$ forced by the atmospheric and CHOOZ results to
the $|U_{e2}|^2$ required for the LMA, LOW or VO solution.
But in the case of the SMA solution, (\ref{ue2-limit})
and (\ref{ue2new}) give
\be
\sin \bar{\Omega} \cos \bar{\Psi} \approx
0.02 \div 0.05 ~~,
\label{cond3}
\ee
which can be satisfied simultaneously with
(\ref{cond1}) and (\ref{cond2}). The region in the
$\bar{\Omega} - \bar{\Psi}$ parameter space that 
satisfies all the constraints
(\ref{cond1}), (\ref{cond2}) and (\ref{cond3})
is shown in Fig.~\ref{param-space}. 
Our scheme thus supports the SMA solution:
if we start with a small $\Omega$ at the high scale
(which is natural in the quark-lepton unified theories),
it does not change much through radiative corrections
(as we have shown in sec.~\ref{mag}), and a small
$\bar{\Omega}$ is retained at the low scale.

\begin{figure}
%\vspace{3.0in}
\epsfig{file=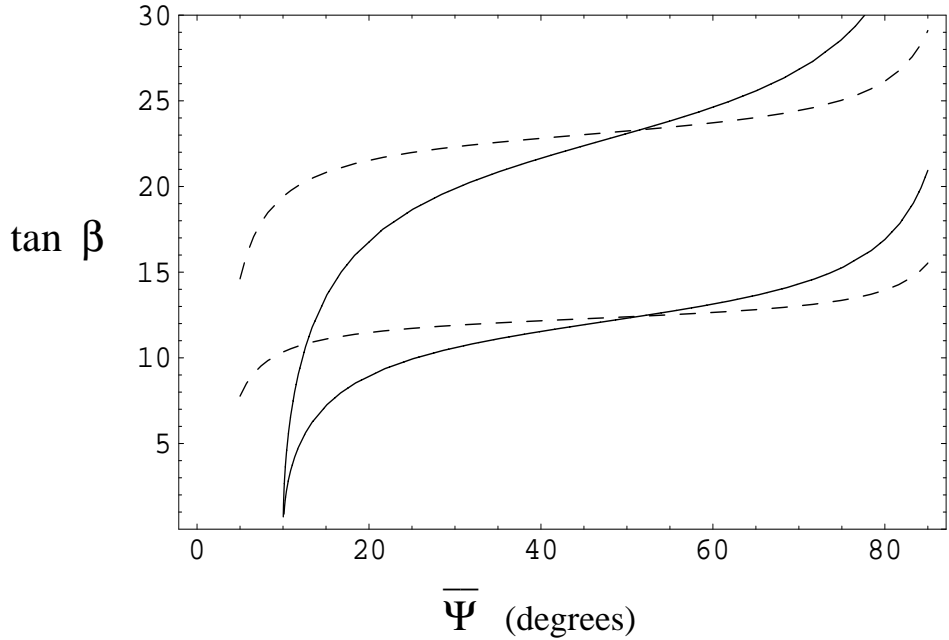,width=5in}
\vspace{0.5in}
\caption{The value of $\tan \beta$ required in MSSM to get a large
$\bar{\Psi}$ from a small $\Psi$. The solid (dashed) curve stands 
for $\Psi = 3^o ~(10^o)$. In each set, the lower (upper) curve
denotes $\Delta m^2(\Lambda)/m^2(\Lambda) = 
2 \cdot 10^{-3} ~(7 \cdot 10^{-3})$
eV$^2$ for the $\nu_2$-$\nu_3$ pair. 
We have chosen $\Lambda/\mu \sim 10^{10}$.}
\label{tanbeta}
\end{figure}

As pointed out in \cite{balaji},
the mechanism of radiative magnification 
does not need any fine-tuning, but is at work
in a range of parameter space for any given model.
As an example of the radiative magnification of $\Psi$, let us
consider MSSM, where the parameter $\tan \beta$ determines
the magnitude of the radiative corrections. The value of 
$\tan \beta$ required to obtain any given magnified value of
$\bar{\Psi}$ is shown in Fig.~\ref{tanbeta}. This indicates
the phenomenologically interesting range of $\tan \beta$
for radiative magnification.

\section{Four neutrino schemes}
\label{4nuscheme}

The features of radiative magnification noted here can be 
used in order to identify the 4$\nu$ mixing 
scenarios in which
the large atmospheric mixing can be naturally
generated. 
Taking into account that the recent atmospheric neutrino results disfavor
($\nu_{\mu}-\nu_s$) oscillations \cite{no-s}, 
the 4$\nu$ solution for all the
anomalies (atmospheric \cite{atm}, solar \cite{solar}
and LSND \cite{lsnd}) 
is essentially of the form
\cite{4nu}
\be
[\nu_e - \nu_s] --- [\nu_\mu - \nu_\tau]~~,
\label{4nu}
\ee
where the $[\nu_e - \nu_s]$ pair 
($\Delta m^2_{14} \approx \Delta m^2_\odot$)
and the
$[\nu_{\mu}-\nu_{\tau}]$ pair 
($\Delta m^2_{23} \approx \Delta m^2_{atm}$)
are separated by $\Delta m^2_{es-\mu\tau} \approx 
\Delta m^2_{LSND}$.
The solar neutrino puzzle is explained by the 
$\nu_e \leftrightarrow \nu_s$
oscillations and the atmospheric data 
are explained by the $\nu_{\mu} \leftrightarrow \nu_{\tau}$
oscillations. A small  $\nu_e-\nu_{\mu}$ mixing then
explains the LSND \cite{lsnd}
observations.

In (\ref{4nu}), the neutrinos can be considered to be
written in the increasing order of masses. With the current
data, it is still possible to change the order 
of neutrinos within a bracket, or the order of the
brackets themselves.
The order within a bracket will not have any influence 
on our conclusions,
so we have only the two independent cases:
(a) $m_{es} < m_{\mu\tau}$ and 
(b) $m_{es} > m_{\mu\tau}$, 
where $m_{es}$ ($m_{\mu \tau}$) denotes the
average mass of the $[\nu_e-\nu_s]$ ($[\nu_\mu - \nu_\tau]$) pair.
In the case (a), $\nu_{\mu}$ and $\nu_{\tau}$ are necessarily
quasi-degenerate: taking $\Delta m^2_{LSND} \sim 1$ eV$^2$
and $\Delta m^2_{atm} \sim 4 \times 10^{-3}$ eV$^2$, we get
the degree of degeneracy ($\frac{\delta m}{m}$) 
for the $\nu_\mu-\nu_\tau$ pair as 
$\frac{\delta m}{m} < 2 \times 10^{-3}$. Then
the $\mu-\tau$ mixing angle $\theta_{\mu \tau}$ 
can be radiatively magnified, as we require for 
the atmospheric neutrino solution.

In the case (b), the neutrinos
$\nu_{\mu}$ and $\nu_{\tau}$ need not be
quasi-degenerate, so the magnitude of
radiative corrections needed to magnify $\theta_{\mu \tau}$ is large. 
Accounting for the large
$\theta_{\mu \tau}$ through radiative magnification is then difficult.
Thus, if radiative magnification is the reason for the large
$\theta_{\mu \tau}$, then the case (a) is favored, {\it i.e.} 
$m_{es} < m_{\mu \tau}$ on the grounds of naturalness.

\section{Conclusions}
\label{concl}

We have shown that, with the parametrization
$U = U_{12}(\Omega) \times U_{13}(\Phi) 
\times U_{23}(\Psi)$
of the lepton mixing matrix, $\bar\Phi=0$, 
$\bar\Psi \approx \pi/4$ and a small $\bar\Omega$
at the low scale
can satisfy all the constraints from the solar, atmospheric
and CHOOZ data. These mixing angles can be generated at
the high scale with $\Phi = 0$ and small $\Omega$ and
$\Psi$ (which is natural in the quark-lepton unified theories),
and magnifying $\Psi$ through radiative corrections while
keeping $\Omega$ and $\Phi$ unaffected.

Let us add a few words on the realization
of this scenario of radiative magnification
in the unified theories. We have not given 
any specific model realization, rather we have pointed out a
class of models that would be successful in generating a large
lepton mixing naturally, starting from a small mixing
at the high scale.
It is not hard to see that such small mixing angle patterns can
emerge at the high scale $\Lambda$ in 
quark-lepton unified theories of type
$SU(2)_L\times
SU(2)_R\times SU(4)_c$ if the right-handed 
neutrino coupling is assumed to
be an identity matrix since the Dirac mass matrix for neutrinos that goes
into the seesaw matrix is then identical to the up-quark mass matrix.
Thus even though our discussion in this paper is completely model
independent, its realization in the context of unified theories is quite
straightforward.
Our work thus demonstrates a way to have a natural solution for 
the neutrino anomalies in the quark-lepton unified theories.

\begin{center}
{\bf Acknowledgements}
\end{center}
We thank WHEPP-6, Chennai, India, where this work was initiated.
The work of RNM is supported by the NSF Grant no. PHY-9802551. The
work of MKP is supported by the project No. 98/37/9/BRNS-cell/731 of 
the Govt. of India. A. D. would like to thank A. De Gouvea for helpful 
discussions. Balaji wishes to thank S. Uma Sankar for useful
clarifications.

\end{document}